\documentclass{elsart}
\usepackage{epsfig,array,amscd}
\usepackage{amsmath,amssymb}
\def\bf{\bfseries}
\begin{document}

\setcounter{page}{1}

\frenchspacing

\sloppy













\begin{flushright}
CERN-PH-EP/2004-048

August 26, 2004
\end{flushright}
\vspace*{10mm}
\begin{frontmatter}

\title{ Measurement of $K^{0}_{e3}$ form factors}
\date{}
\collab{NA48 Collaboration}
\author{A.~Lai},
\author{D.~Marras}
\address{Dipartimento di Fisica dell'Universit\`a e Sezione dell'INFN di Cagliari, \\ I-09100 Cagliari, Italy} 
\author{A.~Bevan},
\author{R.S.~Dosanjh},
\author{T.J.~Gershon},
\author{B.~Hay},
\author{G.E.~Kalmus},
\author{C.~Lazzeroni},
\author{D.J.~Munday},
\author{E.~Olaiya\thanksref{threfRAL}},
\author{M.A.~Parker},
\author{T.O.~White},
\author{S.A.~Wotton}
\address{Cavendish Laboratory, University of Cambridge, Cambridge, CB3~0HE, U.K.\thanksref{thref3}}
\thanks[thref3]{Funded by the U.K.\ Particle Physics and Astronomy Research Council}
\thanks[threfRAL]{Present address: Rutherford Appleton Laboratory, Chilton, Didcot, Oxon, OX11~0QX, U.K.}
% ...

\vspace*{\fill}
\end{frontmatter}

\setcounter{footnote}{0}



\newpage
{\bf Abstract}.

The semileptonic decay of the neutral K meson,  
$K^0_L\rightarrow\pi^{\pm}e^{\mp}\nu$ ($K_{e3}$), was used  
to study  the strangeness-changing weak interaction of hadrons. 
A sample of 5.6 million reconstructed events recorded by the NA48 
experiment
was used to measure the Dalitz plot density.
Admitting all possible Lorentz-covariant couplings, 
the form factors for
vector ($f_+(q^2)$), scalar ($f_S$) and 
tensor ($f_T$) interactions were measured. 
The linear slope of the 
vector form factor $\lambda_+=0.0284\pm 0.0007\pm0.0013$ and values
for the ratios  
$|f_S/f_+(0)|=0.015^{+0.007}_{-0.010}\pm 0.012$ and 
$|f_T/f_+(0)|=0.05^{+0.03}_{-0.04}\pm 0.03$ were obtained.
The values for $f_S$
 and $f_T$ are 
consistent with zero. Assuming only Vector-Axial vector couplings,  
$\lambda_+= 0.0288\pm 0.0004\pm0.0011$   and a good fit consistent with 
pure V-A couplings were obtained.
Alternatively, a fit to a dipole form 
factor
yields a pole mass of $M=859\pm18$ MeV, consistent with the $K^*(892)$  mass.

\section{Introduction}
The study of  semileptonic decays of $K_L$ mesons  gives 
valuable information about the strangeness-changing weak interaction 
 and can be used as a test for possible non-vectorial components
 of the 
weak hadronic interaction and of models of low-energy strong  interactions.

The most general form of the matrix element for $K_{e3}$ decays is given by\cite{Braun} :

\begin{equation}
\label{matr}
 M=\sqrt{1/2} G_F \sin \theta_c \overline{u_{\nu}}(1-\gamma_5)[m_K f_S 
 +~~~~~~~~~~~~~~~~~~~~~~~~~~~~~~~~~~~~~~~~~~~~~~~~~~~~
\end{equation} 
~~~~~~~~~~~~~~~$ (1/2 i)\{(P_K+P_{\pi})_{\lambda} f_+ +(P_K-P_{\pi})_{\lambda} f_-\} \gamma_{\lambda} +i (f_T/m_K) 
 \sigma_{\lambda \tau} 
[P_K]_{\lambda}[P_{\pi}]_{\tau}]u_l$,\\


where $\theta_c$ is the Cabibbo angle, $\overline{u_{\nu}}$ and $u_l$ are
 the lepton 
spinors,
$P_K$ and
$P_{\pi}$ are the kaon and pion four-momenta respectively and  $m_K$ is the kaon
mass. 
The determination of the scalar ($f_S$), 
vector ($f_+$ and $f_-$) and tensor  ($f_T$) form factors 
is based on measurement of the Dalitz plot density which in the kaon
 rest frame has the following form \cite{Chizhov}: 
\begin{equation}
\label{Dalitz_Ke3}
\rho(E^*_{\pi},E^*_{e})\sim a [f_{+}(q^2)]^2 + c [f_S+{1\over m_K} ( E^*_{\nu}-E^*_{e}) f_T]^2,
\end{equation}
where $E^*_i$ is the  energy of the particle $i$ 
in the kaon rest frame,
\newline
$~f_{+}(q^2)=f_{+}(0)(1+\lambda_+
q^2/m_{\pi}^2+\lambda^{\prime}_+q^4/m_{\pi}^4)$,  
\newline
~$~a= m_K(2E^*_{e}E^*_{\nu}-m_K {E_{\pi}}^{\prime})$, 
\newline
~$~c=m^2_K{E_{\pi}}^{\prime}$,
\newline
~$~{E_{\pi}}^{\prime}={(m^2_K+m^2_{\pi})\over 2m_K}-E^*_{\pi}$,
\newline
 $~q^2=(m^2_K+m^2_{\pi}-2m_K E^*_{\pi})$


In the expression for the vector form factor $V$, 
the $f_-$ contribution can be neglected
because it is proportional  to the  electron mass squared. The
dependence of the $f_+$ form factor on the momentum transfer $q^2$ and its value  $f_+(0)$ at
$q^2=0$ are of theoretical interest \cite{Post}, \cite{Cirigliano}, \cite{Becirevic}. 
The $q^2$ dependence is usually assumed to follow 
a pole-dominance formula
$f_+(q^2)=f_+(0) /(1- q^2/M^2)$, where $M$ is the $K^*$ meson  mass, $M=892$ MeV.
This leads in a linear approximation to 
$\lambda_+=m_{\pi}^2/M^2=0.0245$ if one uses  the $\pi^+$ 
mass for $m_{\pi}$. A calculation in chiral perturbation theory to  
order O($p^6$) 
\cite{Post}
predicts a value of $\lambda_+= 0.022$. Other calculations  in chiral perturbation
theory  \cite{Cirigliano}
 and in lattice QCD  \cite{Becirevic} concentrate on the value 
of $f_+(0)$, which is important
for the absolute rate of $K_{e3}$ decay.

Evidence for non-zero scalar and tensor
 form factors in the case of $K^+\rightarrow\pi^{0}e^{+}\nu$ 
has been reported in \cite{Akimenko}.
 Recent measurements 
\cite{Ajinenko}, \cite{Levchenko}  of the charged decay modes have not 
confirmed 
the results of \cite{Akimenko} however, while investigations of
neutral kaon decays have revealed no significant deviation from vector type
interactions 
(\cite{Blumenthal}, \cite{EPJ}). 
The study presented here improves the statistical and systematic 
significance  of
these investigations using a measurement performed with
the NA48 detector in a neutral kaon beam at the CERN SPS.

\section{Experimental setup}
The NA48 detector was designed for a measurement of direct CP
 violation in the $K^0$ system. Here we use data from a 
dedicated run in September 1999
where a  $K_L$ beam was produced by 450 GeV/c protons from the
 CERN SPS incident on a beryllium target.
The decay region is located 
120 m from the $K_L$ target after three 
collimators and sweeping magnets. It is
 contained in an evacuated tube, 90 m long, terminated by a thin
 ($3\cdot 10^{-3} X_0$) kevlar window.

The detector components relevant for this measurement include
 the following:

The  { \bf magnetic spectrometer} is designed to measure the momentum of 
charged particles with high precision. The momentum resolution  is  given by
\begin{equation}
\frac{\sigma(p)}{p} = (0.48 \oplus 0.009 \cdot p) \%
\end{equation}
where $p$ is in $\rm{GeV}/c$. The  spectrometer consists of four drift chambers (DCH), each with 8 planes of sense wires 
oriented along the projections $x$,$u$,$y$,$v$, each
one rotated by 45 degrees with respect to the previous one.
The spatial resolution achieved per projection is 
$\rm{100 \mu m}$ and the time resolution is $\rm{0.7~ns}$.
The volume between 
the chambers is 
filled with helium, near atmospheric pressure. The spectrometer 
magnet is a dipole
with a field integral of 0.85 Tm and is placed after the first two chambers.
The distance between the first and last chamber is 21.8~m.

 The {\bf hodoscope} is placed downstream of the last drift chamber. It
consists of two planes of scintillators segmented in horizontal and vertical
strips and arranged in four quadrants. The signals are used
for a fast coincidence of two charged particles in the trigger. The time
resolution from the hodoscope is 200 ps per track.


The  {\bf electromagnetic calorimeter} (LKr) is a quasi-homogeneous calorimeter based on liquid krypton, 
with tower read out. The 13248 read-out cells have cross sections of 
2 x 2 cm$^2$. 
The electrodes extend from the front to the back
of the detector in a small angle accordion geometry.
The LKr calorimeter measures the energies of the $e^{\pm}$ and $\gamma$ quanta by 
gathering the ionization from their electromagnetic showers.
The energy resolution is :
\begin{equation}
\frac{\sigma(E)}{E} = 
(\frac{3.2}{\sqrt{E}}\oplus\frac{9.0}{E}\oplus0.42) \%
\end{equation}
where $E$ is in GeV, and the time resolution for showers with 
energy between 3 GeV and 100 GeV is $500~ps$.


The {\bf muon veto system} (MUV) consists of three planes of 
scintillator counters, shielded by iron walls  of 80~cm thickness. 
It is  used to reduce the $K_L \rightarrow \pi^{\pm}\mu^{\mp}\nu$
background.


A more detailed description of the NA48 setup can be found elsewhere 
\cite{NA48}. 


\section{Data processing and Monte Carlo simulation}
\subsection{Trigger and data taking}

The trigger was built in two levels. In the first level trigger, the
presence
of at least two hits in the  hodoscope was required.
In  the second level trigger, the drift chamber 
hits were used to 
reconstruct tracks and at least one pair
of tracks was required 
to have a transverse separation below 5~cm 
within 4.5 $K_S$ decay lengths from the
end of the last collimator.
A first level control trigger requiring at least one  hit in 
the  hodoscope in coincidence with at least two tracks segments 
in the first DCH, 
downscaled 20 times, was also implemented for trigger efficiency 
calculations.    

\subsection{Event selection }
The data sample consisted of about 2 TB of data from 100 
million triggers, with approximately equal amounts recorded 
with alternating spectrometer magnet polarities.
The data were 
reconstructed and subjected
to off-line filtering.
The following selection criteria were applied to the 
reconstructed data
to identify $K_{e3}$ decays and to reject background,
keeping in mind the
main backgrounds to $K_{e3}$, which are 
$K_L \rightarrow \pi^{\pm}\mu^{\mp}\nu$ ($K_{\mu3}$) and
$K_L \rightarrow \pi^+\pi^-\pi^0$ ($K_{3\pi}$):     

- Each event was required to contain exactly two tracks, 
of opposite charge,
and a reconstruced vertex in the decay region.
To form a vertex, the
closest distance of approach 
between these tracks had to be less than 3~cm. 
The decay region was defined by
requirements that the vertex had to be between 6 and 34~m from 
the end of the last collimator
and that the transverse distance between the vertex and the
beam axis
had to be less than 2~cm.
These cuts were passed by 35 million events.

- The time difference between the tracks was required to be
less than 6~ns.
To reject muons,
only events with
both tracks inside the detector acceptance and without 
in-time hits in
the MUV system were used.
For the same reason only 
particles with a momentum
larger than 10~GeV ($p_{min}$) were accepted.
In order to 
allow a clear separation of pion and 
electron showers,
we required the distance between the entry points of the  
two tracks at the front face of the LKr Calorimeter ($D_{lkr}$) 
to be larger than 25~cm. As a result 14 million events remained.

\begin{figure}[htb]
\begin{center}
\epsfig{file=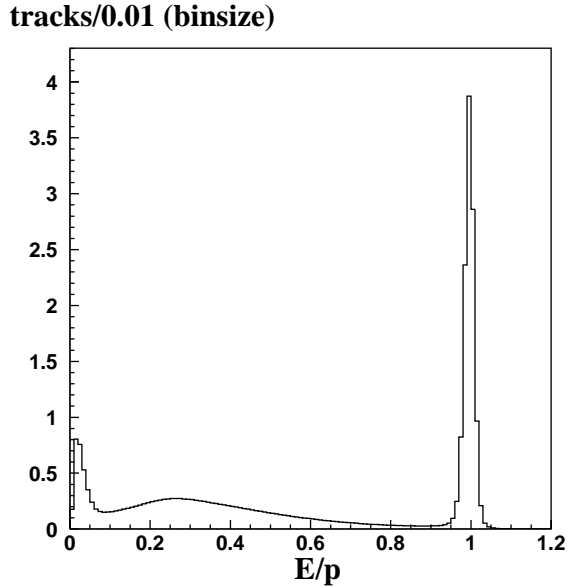,width=8cm,height=8cm}
\end{center}
\caption{Distribution of the ratio of the shower energy $E$
 reconstructed by the LKr and
the momentum $p$ reconstructed by the spectrometer.}\label{eovp}
\end{figure}

- For the identification of electrons and pions, we use the ratio of the
 measured cluster energy, $E$, in the LKr calorimeter associated to a 
track to the momentum, $p$, of this track as measured in the magnetic 
spectrometer. The ratio $E/p$ is shown for all tracks of the 14 million 
events
in fig.\ref{eovp}.
For the selection of $K_{e3}$ events, we require one track to have 
$0.93 < E/p < 1.10$ (electron) and the other track to have 
$E/p < 0.90$ (pion). 11.7 million events were accepted.

- In order to reduce background from $K_{3\pi}$ decays, 
we required the 
quantity 
\begin{equation}
 {P_0^{\prime}}^2=\frac{(m_K^2 -m_{+-}^2 -m_{\pi^0}^2)^2-
  4(m_{+-}^2m_{\pi^0}^2+m_K^2p_{\bot}^2)}{4(p_{\bot}^2+m_{+-}^2)}
\end{equation}
to be less than $-0.004 (\rm{GeV}/c)^2$. 
In the equation above, $p_{\bot}$ is the transverse momentum of the two
 track system  (assumed to consist of two charged pions) 
relative to the  $K_L^0$ flight
direction and   $m_{+-}$ is the
invariant mass of the charged system. 
The variable ${P_0^{\prime}}^2$ peaks at zero if the charged particles
are pions from the decay $K_{3\pi}$.
The cut removes $(98.94\pm0.03)\%$ 
of  $K_{3\pi}$
 decays and $(1.03\pm0.02)\%$ of
$K_{e3}$ decays as estimated with the Monte Carlo simulation (Sect. 3.4). 
After this cut, we were left with 
11.4 million $K_{e3}$ candidate events.

The neutrino momentum in $K_{e3}$ decays
is not known and 
the kinematic reconstruction of the kaon momentum from the measured 
track momenta leads to a two-fold ambiguity
in the
reconstructed kaon momentum. 
%
In order to measure the kaon momentum spectrum, we selected 
events in which both solutions for the kaon momentum 
 lie in the same bin of width 8 GeV. These $4\cdot10^5$ events we call 
"diagonal events".

The last selection criteria were the following two 
requirements:  each solution for the kaon energy 
had to be in the energy range (60,180) GeV and the reconstructed 
electron and pion momenta in the center of mass of the kaon had to
lie in the kinematically allowed region of
the Dalitz plot for $K_{e3}$ decays.

 As a result of this selection,
$5.6 \cdot 10^6$
fully reconstructed  $K_{e3}$ events were selected from the total sample.

\subsection{Background}

The main background to $K_{e3}$ events arises from 
$K_{\mu3}$ and 
 $K_{3\pi}$ decays.
The background from  
$K_{\mu3}$
was reduced by removing events with an in-time muon signal and by 
the $E/p$ requirements. The probability for a pion from $K_{\mu3}$ 
to fake an electron ($E/p>0.93$) 
was measured to be $5.7 \cdot 10^{-3}$.
This was obtained from
a sample of 75000 pion tracks from $K_{e3}$ events 
where the other track had $E/p > 1.02$.
The inefficiency of the MUV system for the 
total exposure
was 
less than 2\%. From this we estimated the number of fake $K_{e3}$ events
from $K_{\mu3}$ to be at most $400\pm 100$ events.
For the suppression of 
background from $K_{3\pi}$  decay, we used the cut 
${P_0^\prime}^2<-0.004(\rm{GeV}/c)^2$ and 
the $E/p$ requirements. 
The number of background events from this source is estimated 
 to be less than 20 events.



\subsection{Monte Carlo Simulation}

The NA48 detector response was simulated using a
GEANT-based Monte Carlo (MC) \cite{NA48}.  The kaon energy 
spectrum in the MC was taken from the kaon energy distribution 
reconstructed using diagonal
$K_{e3}$ events, as defined above (Sect. 3.2).

 For the generation of $K_{e3}$ MC events, the slope
parameter $\lambda_+$ and the scalar and tensor form 
factors were
 set to zero. 
To take into account
the influence of radiative corrections on the
acceptance,  the Photos package was used (\cite{PHOTOS}).
The Dalitz plot density was then modified by the radiative
corrections as calculated by Ginsberg \cite{Ginsberg}.
This modification was made by randomly rejecting events 
according to the correct 
shape of the Dalitz plot.    
From 45 million generated MC events, 7.7 million were selected
  by the criteria in Sect. 3.2, corresponding
to an  acceptance of about 17\%.


\begin{figure}[h]
\begin{center}
\epsfig{file=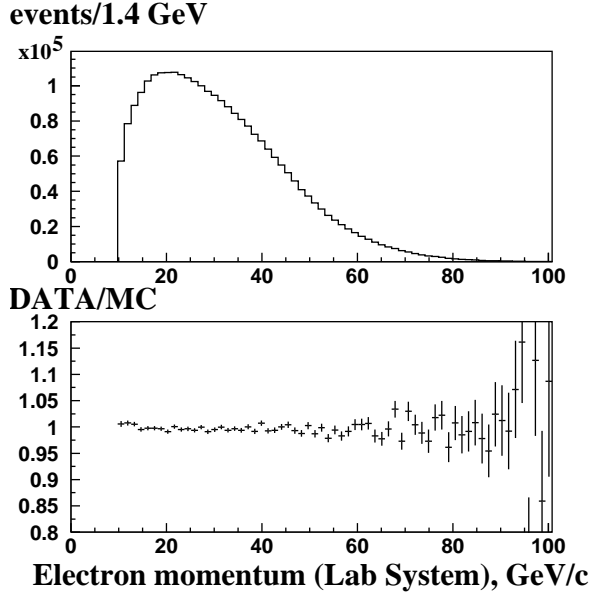,width=8.0cm,height=8cm}
\end{center}
\caption{Reconstructed electron momentum in the 
Laboratory System; data 
distribution and  ratio 
data/MC}\label{p_el}
\end{figure}

\begin{figure}[h]
\begin{center}
\epsfig{file=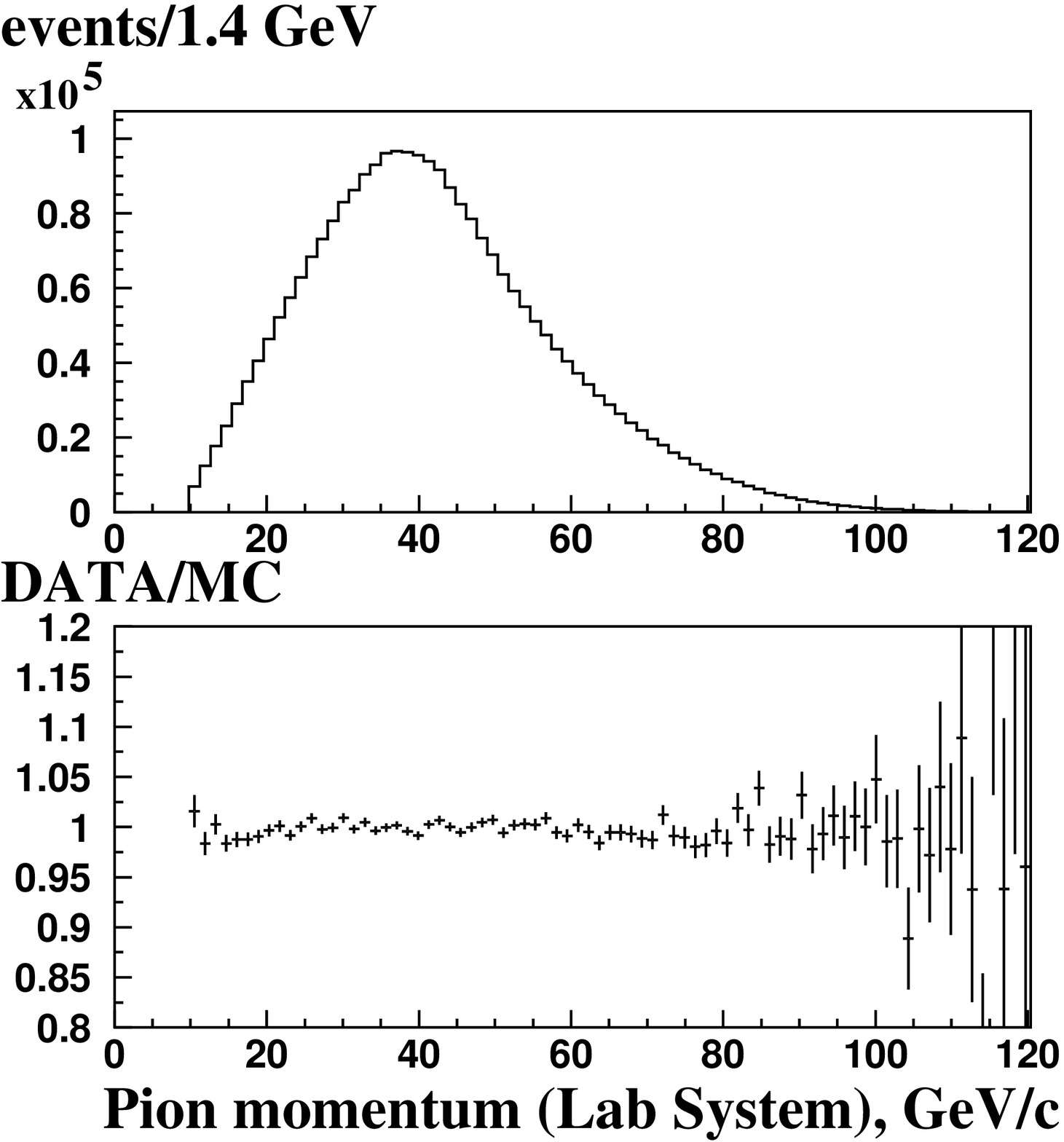,width=8.0cm,height=8cm}
\end{center}
\caption{Reconstructed pion momentum in the 
Laboratory System; data 
distribution and  ratio 
data/MC}\label{p_pi}
\end{figure}

\begin{figure}[h]
\begin{center}
\epsfig{file=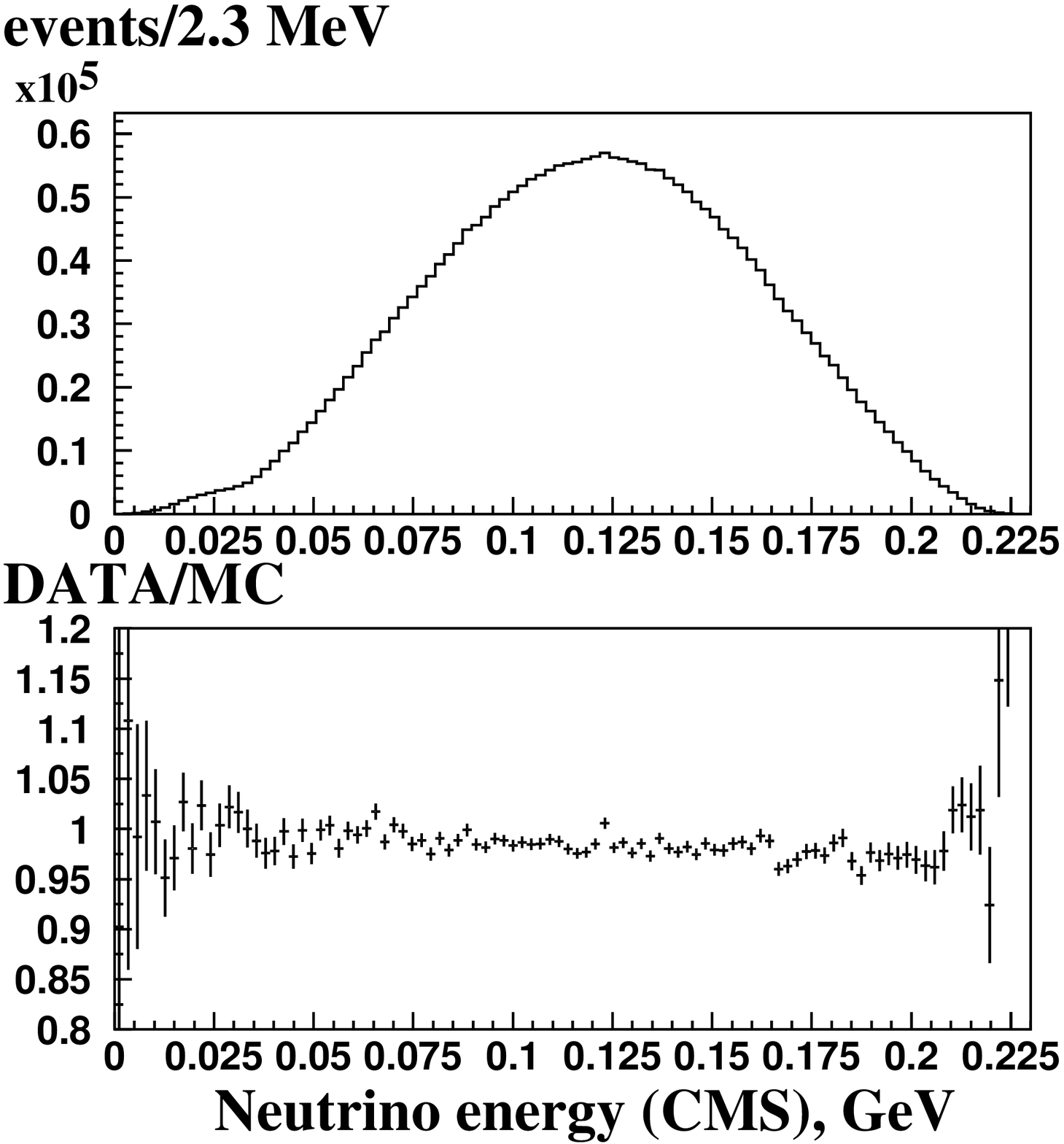,width=8.0cm,height=8cm}
\end{center}
\caption{Reconstructed neutrino energy $E^*_{\nu}$ in the 
Center of Mass System;  
data distribution and  ratio 
data/MC}\label{e_nu}
\end{figure}

Distributions for accepted Monte Carlo events weighted with a 
form factor slope of $\lambda_+=0.029$ are shown in 
figs. \ref{p_el}, \ref{p_pi} and  \ref{e_nu} together with data,
for the negative magnet polarity.
As an  example,
the distributions of charged particle momenta in the 
Laboratory System (figs. \ref{p_el}, \ref{p_pi}) and of the neutrino 
energy $E^*_{\nu}$ in the centre-of-mass system, CMS,
 (fig. \ref{e_nu}) are presented.
The quantity $E^*_{\nu}$ is unambiguously defined.
 The upper part of the figures shows distributions of data, 
while the lower part shows the ratio 
 of the experimental and MC spectra. Good agreement is seen for the 
Monte Carlo simulations with
a form factor slope $\lambda_+=0.029$. 




\section{Data analysis}

\subsection{Dalitz plot analysis}
%

For the Dalitz plot analysis, 
selected $K_{e3}$ events were binned in 
a 19x19x19 three-dimensional array 
$N(E^*_{\nu},q^2_1/m^2_\pi,q^2_2/m^2_\pi)$,
 where the indices 1 and 2 stand for the
 two possible solutions for the reconstructed kaon 
momentum and the bin sizes are 12 MeV for $E^*_{\nu}$ and 
0.35 for the $q^2/m^2_\pi$ bins. Two-dimensional arrays  
$N(q^2_1/m^2_\pi,q^2_2/m^2_\pi)$ 
were used when pure V-A couplings were assumed. 

To derive the values of the form factors we used a maximum
log-likelihood method \cite{B_Ch}
based on the following likelihood function which takes into 
account the finite size of the MC sample:

\begin{equation}
\label{log}
\rm{ln}\it{L}=-\rm{2}[\sum_{i}(\it{d_i} 
\ln \it{f_i}-f_i)+\sum_{i}(a_{i}\ln A_{i}-A_{i})],
\end{equation}
 
Here, $d_i$ is the number of data events in the $i$'th bin,
 $f_i$ is  the predicted number of events,
$a_i$ is the actual number of MC events and
$A_i$ is the expected number of MC events. 
The predicted number of events,  $f_i$, is a function of the 
three form factors parameters $\lambda_+$, $f_S$, $f_T$, 
and is obtained by weighting the MC expectation $A_i$ for a flat
Dalitz plot with the form factors according to 
Eqn.\ref{Dalitz_Ke3}. 
This procedure provides a weight $h_i$ 
for each bin.
The MC expectation $A_i$ for the bin $i$ is obtained from the 
relation
$A_i=(d_i+a_i)/(1+h_i)$, which is valid at the maximum of the 
likelihood
function \cite{B_Ch}, \cite{SS}. 
The
predicted number $f_i$ to be compared with the  data 
$d_i$ is then $f_i=A_i\cdot h_i$.


In order to test the goodness of fit, the variable

\begin{equation}
\label{chi}
\chi^2_{gof}=
[\sum_{i}{(f_i-d_i+d_i \ln\frac{d_i}{f_i}  )}+
\sum_{i}{(A_{i}-a_{i}+a_{i} \ln\frac{a_{i}}{A_{i}})}] 
\end{equation}

was used \cite{EPJ}, \cite{Baker}.
The minimization of the
log-likelihood function, using MINUIT \cite{MINUIT},
gives the values of the fitted  form factors.
     



\subsection{Results}
\begin{figure}[t]
\begin{center}
\epsfig{file=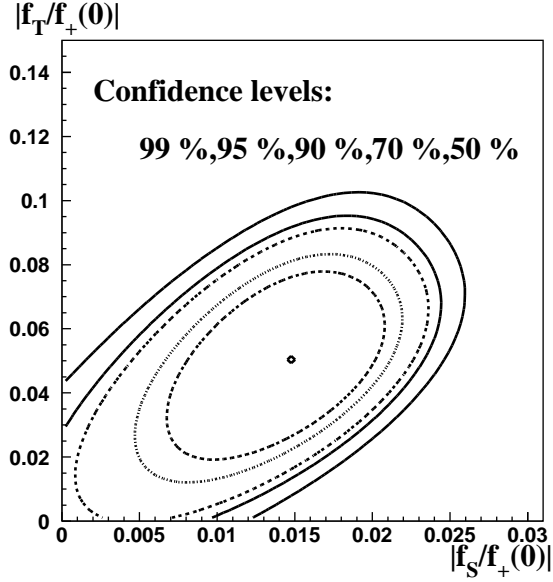,width=8.0cm,height=8cm}
\end{center}
\caption{Confidence level contours in the 
($|{f_S / f_+(0)}|$,$|{f_T / f_+(0)}|$) plane at 
$\lambda_+=0.0284$. The outermost contour corresponds to 
99\% C.L., the innermost to 50\% C.L. Only statistical errors are
considered.
}\label{fs_ft}
\end{figure}

\begin{figure}[]
\begin{center}
\epsfig{file=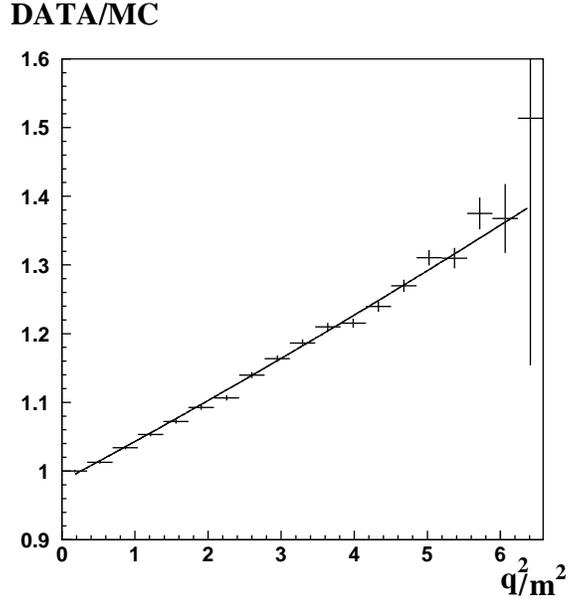,width=8.0cm,height=8cm}
\end{center}
\caption{Ratio of ``diagonal'' data and MC events 
(with $\lambda_+=0$). The vertical axis corresponds to 
$[f_+(q^2)/f_+(0)]^2$. The curve corresponds to 
$\lambda_+=0.0288$}\label{q2_sum}
\end{figure}

Linear fits.

We analysed the Dalitz plot under two hypotheses:  either assuming
pure 
V-A interaction or admitting additional scalar and tensor 
interactions. 
The results for the form factors obtained, assuming
a linear $q^2$ dependence, are presented in Table 1.
The statistical errors
have contributions from fluctuations of the data and of the MC events,
as explained in sect.4.1.


\begin{table}[ht]
\begin{center}
\mbox{
\begin{tabular}{||l|l|l||}
\hline
Form factor     &  value and stat. err.     & $\chi^2$/DOF\\
\hline
   $\lambda_+$       & $ 0.0284 \pm 0.0007$ & 3010/2915\\
\hline
  $|{f_S / f_+(0)}|$  &  $0.015 ^{+0.007}_{-0.010}$    
   &3010/2915\\
\hline
 $|{f_T / f_+(0)}|$  &  $0.05\pm ^{+0.03}_{-0.04}$    
   &3010/2915\\
\hline
\hline
\hline
      $\lambda_+ $      &$ 0.0288\pm 0.0004$ & 394/302\\
\hline
\hline
      $\lambda_+ $      &$ 0.0280\pm 0.0019$ & 386/301\\
\hline
      $\lambda^{\prime}_+ $      &$ 0.0002\pm 0.0004$ & 386/301\\
\hline
\end{tabular}
}
\end{center}
\caption{Results from the fits: 3-form factors case and 1-form factor
case (last rows)}
\end{table}

The confidence level contours in the  ($|{f_S / f_+(0)}|$,$|{f_T
/ f_+(0)}|$) plane for $\lambda_+=0.0284$ are plotted in fig   
\ref{fs_ft}. The results are consistent with both $f_S$
and $f_T$ being zero at 10\% C.L. taking only statistical 
errors into account.

To visualize the main feature of the two-dimensional fit we 
display in
fig. \ref{q2_sum} the ratio of ``diagonal'' data and MC 
events (with $\lambda_+=0$). ``Diagonal'' here means events 
for which both solutions 
for $q^2$ lie in one $q^2/m_{\pi}^2$-bin or they lie 
in neighbouring bins.  
The quantity plotted
corresponds to the square of the 
$q^2$ variation of the vector form factor,
$[f_+(q^2)/f_+(0)]^2=[1+\lambda_+ q^2/m_{\pi}^2]^2$.
The data are seen to be consistent with a linear 
increase of  $f_+(q^2)$ with increasing $q^2$.

Our investigation of possible systematic errors showed that the
 biggest uncertainty for $\lambda_+$ comes from the 
 uncertainty in the
kaon momentum spectrum. In order to
 determine the influence of this factor,
 the kaon momentum spectra derived
from reconstructed   $K \rightarrow \pi^+\pi^-$
and  $K \rightarrow \pi^+\pi^-\pi^0$ decays were implemented 
in the MC.
The differences in the fitted form factor values using 
either of these
spectra compared to the spectrum from diagonal $K_{e3}$ 
events was
 taken as a systematic error.
This error amounts to
0.0007, and is 7 times larger than claimed in ref.\cite{KTeV}.
We studied also the systematic errors induced by the various 
$K_{e3}$ selection cuts, 
by the momentum and energy calibrations,
Dalitz plot distributions bin width, 
inefficiency of the muon veto system and the trigger, 
possible detector asymmetries
and influence of  accidental particles.
 
The systematic uncertainties connected with our knowledge of the
  detector acceptance were evaluated by variations of the selection cuts
  in between
values
which changed the number of accepted events by
up to  20\%. 
The largest fluctuations in the form factor values
were taken as systematic errors.


Possible shifts in the measurement of $p$ and $E$ were
investigated by comparing the value of the reconstructed kaon
 mass from
$K_{3\pi}$ and $K_{2\pi}$ decays with the world average 
for the kaon mass and by comparing
the central value of the $E/p$ distribution of electrons with unity.
These led to a rescaling of $p$ and
$E$ by $1\cdot 10^{-3}$ and $4\cdot 10^{-3}$ respectively.
The influence
on the 
form factor values was studied by scaling systematically  $p$ and
$E$ by the quoted amounts.

The inefficiency of the MUV during this run was measured to
be below 2 \%. In order to investigate the influence of 
the MUV inefficiency we
introduced an extra inefficiency of 2 \% in the 
experimental data and compared
the results. The difference was taken as a systematic error.

The trigger inefficiency  was evaluated by using
 the control trigger and was found to be $(1.9\pm0.1)\%$. 
The influence of this source of systematic uncertainty on the 
 results
was estimated as follows. 
The results obtained with the main trigger data for which the control
trigger was set as well were compared to the results obtained with the
control trigger data which are strongly correlated. The differences are
due to the 2\% events escaping the main trigger, and indicate the
uncertainty due to this inefficiency.
%


To estimate the possible influence of accidental particles we used
the Overlay MC technique, applied earlier in the 
$\epsilon^{\prime}/\epsilon$ analysis \cite{NA48}, in which  
randomly recorded experimental
signals  
with rate proportional to the beam intensity
were mixed with each MC event.  
The uncertainty  
due to this effect  was estimated by comparing results obtained 
with the standard MC 
with results obtained with the Overlay MC.

We also varied the number of bins (from 10 to 25 in each axis) in the
distributions $N(E_{\nu},q^2_1,q^2_2)$ and $N(q^2_1,q^2_2)$. The
change in the values of the fit parameters was considered as
a systematic error.

The individual systematic uncertainties, and the total
systematic error obtained  by combining the individual
errors in quadrature, are summarized in table 2.


\begin{table}[tp]
\begin{center}
\mbox{
\begin{tabular}{||l|l|l|l||l||}
\hline
Source    &  $\triangle \lambda_+ $    & 
$\triangle |{f_S \over f_+(0)}|$ &$\triangle |{f_T \over f_+(0)}|$ &
$\triangle \lambda_+$ only\\
\hline
$K_L$ spectrum& $\pm 0.00080$ &
$\pm 0.001$ &$\pm 0.005$ &$\pm 0.00070$ \\
Geom. accept. &$ \pm 0.00050$ &
 $\pm 0.007$&$\pm 0.015$ &$ \pm 0.00040$\\
$p_{min}$&$\pm 0.00025$   &
$\pm 0.004$&$\pm 0.010$& $\pm 0.00015$\\ 
$D_{lkr}$&$\pm 0.00045$ &
$\pm 0.004$ & $\pm 0.005$&$\pm 0.00025$\\
$E/p$&$ \pm 0.00035$   &
$\pm 0.002$&$\pm 0.010$&$\pm 0.00035$\\
${P_0^{\prime}}^2$   &$\pm 0.00020$& 
$\pm 0.003$&$\pm 0.005$& $\pm 0.00010$\\
$E$,$p$ scaling& $\pm 0.00020$ &
$\pm 0.001$ &$\pm 0.005$ &$\pm 0.00020$\\
MUV ineff. &$\pm 0.00020$   &
$\pm 0.002$&$\pm 0.005$ &$\pm 0.00020$\\
Trigger ineff.&$\pm 0.00015$ & $\pm 0.002$
 &  $\pm 0.005$&$\pm 0.00025$ \\
accidentals& $\pm 0.00030$&  $\pm 0.001$
 & $\pm 0.005$& $\pm 0.00025$\\
bin width &$\pm 0.00040$&
$\pm 0.005$&$\pm 0.010$& $\pm 0.00010$\\
\hline
TOTAL &$\pm 0.0013$&
$\pm 0.012$&$\pm 0.03$& $\pm 0.0011$\\
\hline
\end{tabular}
}
\end{center}
\caption{Systematic uncertainties: 3-form factors case and 1-form
factor case (right column)}
\end{table}

As a cross check, correlations between magnetic field polarities, 
positive and negative tracks, 
particles and anti-particles, 
geometrical positions of the tracks
 were explored. 
We did not find significant
changes in results coming from these sources. 
The effect
of the uncertainties of the radiative corrections 
\cite{Ginsberg} was found to have
a negligible effect on the results.

Quadratic and dipole fits.


As an alternative to these linear slope fits, we also investigated two other
forms of the $q^2$ dependence: a quadratic form, admitting a finite 
$\lambda^{\prime}_+$
as a free parameter, and a pole-dominance form factor of the form given in
Sect.1, with the mass of the vector meson pole mass $M_V$ as the only free
parameter. The quadratic fit gave the values of the two slope coefficients,
$\lambda_+=(28.0\pm1.9\pm1.5)\cdot10^{-3}$ and 
$\lambda^{\prime}_+=(0.2\pm0.4\pm0.2)\cdot10^{-3}$, 
consistent with no quadratic
term but also consistent with a dipole form factor where the quadratic term of
the Taylor expansion corresponds to 
$\lambda^{\prime}_+=\lambda_+^2= 0.7 \cdot 10^{-3}$. 
The dipole fit with the
vector meson mass as free parameter gave $M_V= 859\pm18$ MeV, fully 
consistent with the mass of the $K^*(892)$ meson.

\section{Conclusions}

%

Using 5.6 million $K_{e3}$ events recorded by the NA48 detector, we tested 
the
validity of the V-A weak interaction and measured the 
$q^2$ dependence of the
vector form factor with high precision.
Admitting scalar and tensor couplings in addition to V-A couplings , we obtain
the following form factor values :



\begin{center}
$|\frac{f_{S}}{f_{+}(0)}| = 0.015 ^{+0.007}_{-0.010} \pm 0.012$   

$|\frac{f_{T}}{f_{+}(0)}| = 0.05  ^{+0.03}_{-0.04} \pm 0.03$ 

\end{center}

and the linear slope in the vector form factor 
$\lambda_{+}=(28.4\pm0.7 \pm 1.3)\cdot 10^{-3}$.
The $90\%$ C.L. upper limits on the scalar and tensor form factors are
$|\frac{f_{S}}{f_{+}(0)}|<0.041$ and  $|\frac{f_{T}}{f_{+}(0)}|<0.12$.
We therefore see no evidence for scalar or tensor couplings, not
supporting the results of ref. \cite{Akimenko}.

\begin{figure}[t]
\begin{center}
\epsfig{file=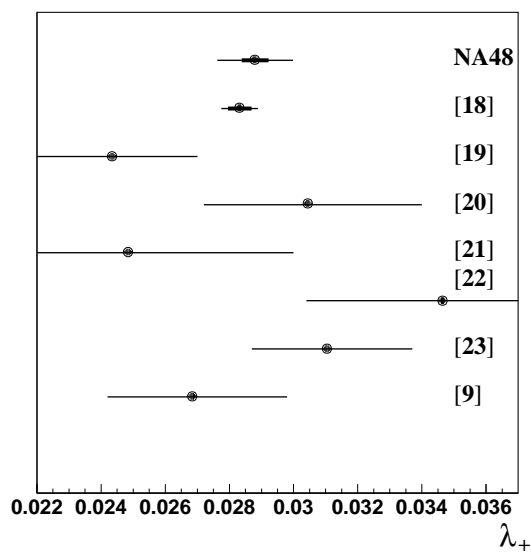,width=8.0cm,height=8cm}
\end{center}
\caption{Experimental results for the slope $\lambda_+$ of the vector
form factor. The bold error bars are the statistical errors, while the thin
line indicates the total uncertainty including the systematic error.}\label{rescompar}
\end{figure}

Assuming only V-A currents and a linear $q^2$ dependence the 
$\lambda_+$ value is
$\lambda_{+}=(28.8\pm0.4\pm1.1) \cdot10^{-3}$.
If in addition we admit quadratic terms , the results are 
$\lambda_+=(28.0\pm1.9\pm1.5)\cdot10^{-3}$ and
$\lambda^{\prime}_+ =(0.2\pm0.4\pm0.2)\cdot10^{-3}$.
We
therefore find no evidence for a quadratic 
term $\lambda^{\prime}_+$ different from zero. However,
the size of a quadratic term is consistent with a Taylor expansion of a
pole-dominance form factor. Such a dipole form factor is in good agreement with
the data, with a pole mass of $859\pm18$ MeV.


Fig.  \ref{rescompar} shows the linear slope.



Our results agree well with the V-A coupling of the weak
interaction and with a pole form of the form factor, in agreement with the
linear approximations tested in former experiments
 \cite{Blumenthal}, \cite{KTeV}, \cite{CPLEAR},
\cite{Birulev}, \cite{Engler}, \cite{Hill}, \cite{Gjesdal}.
Our result for the quadratic term in the vector form factor is at variance
with the result of ref.
\cite{KTeV}
 though it agrees 
with the
pole model form factor.






\end{document}